\documentclass[conference]{IEEEtran}
\IEEEoverridecommandlockouts
\usepackage{cite}
\usepackage{amsmath,amssymb,amsfonts}
\usepackage{algorithmic}
\usepackage{graphicx}
\usepackage{textcomp}
\usepackage{xcolor}
\usepackage{siunitx}
\usepackage{comment}
\usepackage{stfloats}
\def\BibTeX{{\rm B\kern-.05em{\sc i\kern-.025em b}\kern-.08em
    T\kern-.1667em\lower.7ex\hbox{E}\kern-.125emX}}

\begin{document}

\title{A Wireless Self-Calibrating Ultrasound Microphone Array with Sub-Microsecond Synchronization}

\author{
\IEEEauthorblockN{Dennis Laurijssen\IEEEauthorrefmark{1}\IEEEauthorrefmark{2}, 
Rens Baeyens\IEEEauthorrefmark{1}\IEEEauthorrefmark{2}, 
Walter Daems\IEEEauthorrefmark{1}\IEEEauthorrefmark{2}, 
Jan Steckel\IEEEauthorrefmark{1}\IEEEauthorrefmark{2}}
 \IEEEauthorblockA{\IEEEauthorrefmark{1}Cosys-Lab, Faculty of Applied Engineering, University of Antwerp, Antwerp, Belgium}
 \IEEEauthorblockA{\IEEEauthorrefmark{2}Flanders Make Strategic Research Centre, Lommel, Belgium\\
 \IEEEauthorrefmark{1}dennis.laurijssen@uantwerpen.be}
}

\maketitle

\begin{abstract}
We present a novel system architecture for a distributed wireless, self-calibrating ultrasound microphone network for synchronized in-air acoustic sensing. Once deployed the embedded nodes determine their position in the environment using the infrared optical tracking system found in the HTC Vive Lighthouses. After self-calibration, the nodes start sampling the ultrasound microphone while embedding a synchronization signal in the data which is established using a wireless Sub-1GHz RF link. Data transmission is handled via the Wi-Fi 6 radio that is embedded in the nodes' SoC, decoupling synchronization from payload transport. A prototype system with a limited amount of network nodes was used to verify the proposed distributed microphone array's wireless data acquisition and synchronization capabilities. This architecture lays the groundwork for scalable, deployable ultrasound arrays for sound source localization applications in bio-acoustic research and industrial acoustic monitoring.

\end{abstract}


\section{Introduction}
Microphone arrays capable of recording high-frequency airborne ultrasound have become instrumental in various applications, including acoustic imaging using beamforming~\cite{Izquierdo2017, Izquierdo2020, Izquierdo2024, Haugwitz2024, Maier2021, 10210599} and the study of bio-acoustics~\cite{okamoto2025chirparray, MARTINEZ2023109618}. Achieving higher spatial resolution requires more microphones placed further apart, while higher temporal resolution demands precise synchronization and high-bandwidth data acquisition. These requirements motivate the need for distributed and scalable sensor networks.

Traditionally, synchronized recordings in large-scale arrays are achieved through centralized, wired setups~\cite{Derogarian2014}. These systems rely on high-end microphones coupled to multi-channel data acquisition units, where synchronization and data integrity are maintained through physical connections. While such solutions offer robust performance, they are inherently limited in terms of scalability, cost, and deployment flexibility. Their cumbersome nature becomes particularly evident in field experiments or applications where sensor mobility and rapid reconfiguration are required.

Some recent efforts have aimed to overcome these constraints by adopting distributed embedded systems~\cite{Djenouri2016,Sondej2024,Biagetti2025,Li2025}. A representative example is the system described in~\cite{verreycken2021bio}, where the authors utilize a network of embedded (BeagleBone Black) devices with custom ultrasound receivers. However, such approaches still face limitations regarding long (fixed) cable lengths, centralized control, and the manual calibration of node positions.

One of the primary practical challenges in these large-scale microphone arrays is spatial calibration—determining the exact position of each sensor node in a common world coordinate system~\cite{Peng2014, Jameel2017, Jiang2020, Ahmad2024}. Accurate localization of the microphones is essential for any spatial audio processing task. Existing solutions often rely on manual measurements or labor-intensive calibration procedures involving external tracking systems or time-consuming acoustic localization routines.

To address these limitations, we propose a novel system architecture for a distributed, synchronized ultrasound microphone network that is self-calibrating and wirelessly deployed. Each node in the proposed network is a fully embedded device capable of synchronized ultrasound acquisition and automatic spatial localization using two HTC Vive Lighthouse base stations.. This enables fast, plug-and-play deployment of large-scale microphone arrays, greatly reducing setup complexity and removing the need for cumbersome wiring or manual calibration.

\section{System Architecture}
The proposed system is a fully embedded, wireless ultrasound microphone network designed for synchronized in-air acoustic sensing. Each node operates autonomously and is capable of recording ultrasound data, synchronizing in time with other nodes, and estimating its spatial position in a shared coordinate system. The architecture decouples synchronization and data transmission across two radio interfaces: a Sub-1\si{\giga\hertz} (868~\si{\mega\hertz}) link for synchronization and Wi-Fi 6 (at 2.4~\si{\giga\hertz}) for high-speed data transfer.

Nodes are deployed without any physical interconnection and form a scalable, flexible topology. A dedicated synchronization node continuously transmits pseudo-random packets over the Sub-1\si{\giga\hertz} band. Each sensing node receives these packets using a Sub-1\si{\giga\hertz} transceiver and embeds synchronization events directly into the sampled data stream. Meanwhile, Wi-Fi 6 provides a reliable wireless data link to a central server. This dual-band approach enables synchronized high-throughput data acquisition.

\subsection{Embedded Hardware}
\begin{figure}
    \centering
    \includegraphics[width=1\linewidth]{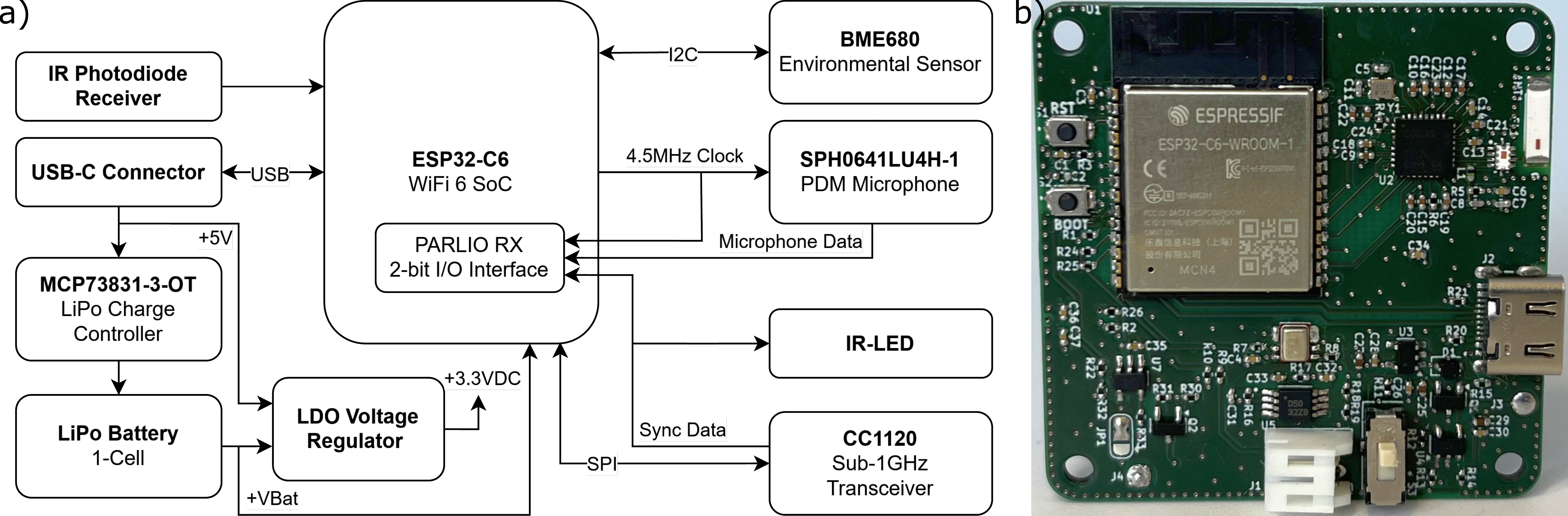}
    \caption{On the left the schematic overview of the hardware architecture of the developed measurement node with an image of the device PCB measuring 45\si{\milli\meter} by 45\si{\milli\meter} on the right.}
    \label{fig:system_architecture}
\end{figure}

Each measurement node is based on a custom 45~\si{\milli\meter} × 45~\si{\milli\meter} PCB, integrating all components required for synchronized ultrasound acquisition and self-calibration. A schematic representation of the node can be seen in figure~\ref{fig:system_architecture}. At its core is the ESP32-C6 SoC, which provides Wi-Fi 6 connectivity and manages data acquisition and peripheral control via its RISC-V core.
For synchronization, the node includes a CC1120 transceiver operating in the 868~\si{\mega\hertz} ISM band, receiving timing packets from a dedicated beacon. Ultrasound sensing is handled by a SPH0641LU4H-1 digital MEMS microphone, providing high-frequency 1-bit PDM output, sampled and timestamped in firmware. Optical self-calibration is supported via a VBP104S PIN photodiode, amplified through a transimpedance and gain stage to detect IR light pulses and laser sweeps from Lighthouse base stations. A BME680 sensor measures temperature, humidity, and pressure to improve local sound speed estimation.
Power is supplied via a single-cell LiPo battery or USB Type-C, with integrated charging and voltage monitoring. An onboard IR LED, connected to the synchronization logic, allows data alignment with external video recordings. This compact, self-contained node design enables scalable deployment of synchronized, wireless ultrasound sensors with minimal setup effort.

\subsection{Self-calibration Methodology}
To eliminate the need for manual spatial calibration of the distributed microphone nodes, the system incorporates an optical self-localization technique based on the HTC Vive Lighthouse (V1) tracking system. This method enables the autonomous determination of each node's 3D-position within a predefined world coordinate system similar to previous work~\cite{laurijssen2017six, laurijssen2019synchronous}.

Two Lighthouse base stations are placed at fixed, known locations in the environment and serve as the optical reference frame. These devices emit a repeating sequence of infrared light pulses followed by horizontal and vertical IR laser sweeps with a fixed rotation speed. 

\begin{figure}
    \centering
    \includegraphics[width=0.95\linewidth]{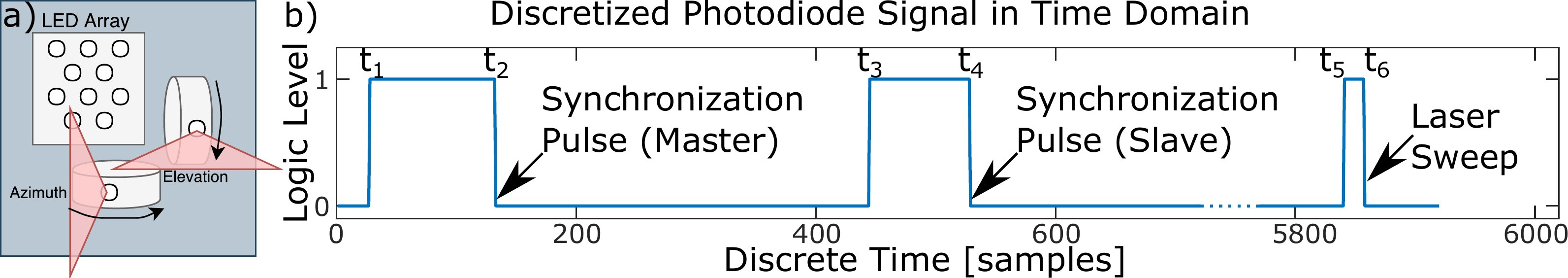}
    \caption{(a) Schematic of a Lighthouse base station. (b) Example photodiode waveform showing synchronization flashes and laser sweep transitions.}
    \label{fig:vivedata}
\end{figure}

Each node is equipped with a photodiode receiver hardware and corresponding analog signal chain. This allows detecting the precise timing of the IR signals emitted by both lighthouses by tracking the microsecond-resolution timestamps of the rising and falling edges of the photodiode output as shown in Fig.~\ref{fig:vivedata}. These edges correspond to the initial synchronization flash and the subsequent laser sweep transitions, enabling accurate angular measurements.

Upon startup, each node passively observes the incoming optical signals and determines the azimuth and elevation angles of both lighthouses by calculating the time offset between the sync pulse and the respective horizontal and vertical sweeps.
Assuming a known node orientation, these angular observations are used in a triangulation algorithm to compute the position of the node relative to the two fixed Lighthouse positions.

Because the optical tracking is performed locally and independently by each node, the system supports asynchronous self-calibration and requires no wired infrastructure or external measurement equipment. This enables rapid and repeatable deployment in arbitrary environments while maintaining high spatial accuracy.

\subsection{Synchronization Methodology}

The system implements a wireless synchronization strategy based on our previous work~\cite{10.1242/jeb.173724, baeyens2024synchronisation}, in which a pseudo-random binary sequence (PRBS) is embedded directly into the data stream for aligning recordings across multiple nodes. This removes the need for wired synchronization or shared clocks. 
A dedicated Sub-1\si{\giga\hertz} transmitter continuously emits packets with randomized payload lengths and inter-packet intervals, producing a unique temporal pattern that functions as a pseudo-random sequence observable by all nodes. Each node includes a CC1120 receiver, which parses these packets using its built-in protocol engine. The start and end of each packet reception toggle a dedicated GPIO pin, generating a square wave with a pseudo-random period that mimics the transmitted sequence.

The node’s SoC samples both the 1-bit PDM microphone data and the CC1120 sync GPIO using the dual-channel PARLIO RX interface clocked at 4.5~\si{\mega\hertz}, ensuring tight temporal alignment of ultrasound and sync signals. After acquisition, synchronization traces are extracted and cross-correlated to estimate inter-node time offsets with sub-microsecond precision. Additionally, the GPIO is connected to an on-board IR LED that emits flashes synchronized to the RF signal, enabling alignment with external video recordings~\cite{10833678}.

\section{Experimental Setup and Results}
To validate the system, a prototype setup with three embedded nodes was assembled, reflecting the early-stage development phase with only a few hand-soldered devices. A Wi-Fi 6 router provided a dedicated 2.4~\si{\giga\hertz} network for wireless data transmission, where the throughput per node was measured to be approximately 17~\si{\mega\bit\per\second} using iperf2. All nodes connected via their ESP32-C6 radios, while a host computer running a custom TCP server received the data streams.

Two nodes performed full data acquisition, capturing PDM ultrasound data and sampling the synchronization signal via the PARLIO RX interface. The third node acted as a synchronization beacon, continuously transmitting pseudo-random packets on the 868~\si{\mega\hertz} ISM band using a CC1120 transceiver. Each acquisition node received and recorded these synchronization events with sub-microsecond accuracy, embedding them directly into the ultrasound data stream.

This compact setup effectively demonstrates synchronized multi-node acquisition, integrated RF-based synchronization, and high-throughput wireless data offloading.
\begin{figure*}[t]
\centering
    \includegraphics[width=0.95\linewidth]{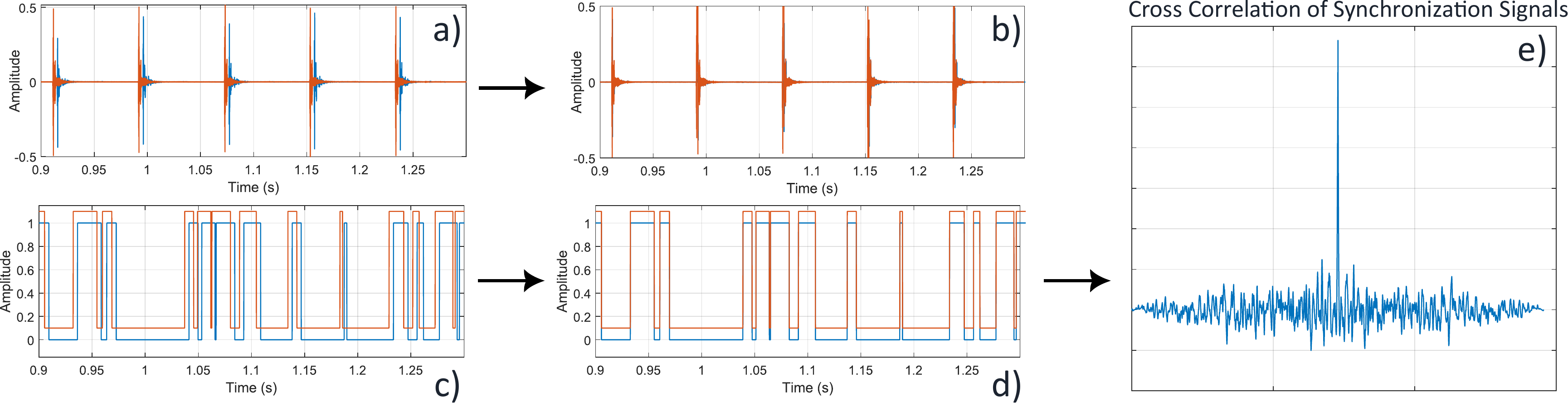}
    \caption{The two plots a) and c) respectively show the recorded microphone and sub-1GHz PRBS sync data of two nodes received over TCP before synchronization. Whereas plots b) and d) show the synchronized data after making using of the cross-correlation, shown in e), of the PRBS sync signals to determine the time offset.}
    \label{fig:enter-label}
\end{figure*}

\subsection{Preliminary Results}

\subsubsection{Optical Localization Calibration}
The angular precision of the Lighthouse-based decoding method was evaluated by comparing computed azimuth and elevation angles against ground truth measurements from a Qualisys motion capture system. Across multiple trials, the standard deviation of the angular error remained low: 0.08° for azimuth and 0.24° for elevation (LH1), and 0.18° for azimuth and 0.39° for elevation (LH2). These results demonstrate a high degree of repeatability, with most errors attributable to systematic offsets rather than random variation.

\subsubsection{Synchronization Accuracy}
To evaluate inter-node synchronization performance, the embedded pseudo-random synchronization signals recorded in each data stream were extracted and cross-correlated. The measured relative time difference between the two acquisition nodes was consistently below 1~\si{\micro\second} across multiple acquisitions, while other reported approaches typically exhibit synchronization errors ranging from several microseconds to milliseconds~\cite{Sondej2024}. This confirms that the CC1120-based synchronization mechanism provides sufficient temporal resolution for typical ultrasound signal processing tasks, such as beamforming or source localization.

\subsubsection{Power Consumption and Battery Life}
Each node was operated from a single-cell 3.7~\si{\volt} 1400~\si{\milli\ampere\hour} LiPo battery. During continuous data acquisition and Wi-Fi transmission, the maximum current draw per node during data acquisition and TCP data transmission was measured at approximately 245~\si{\milli\ampere}, resulting in a total runtime of approximately 5.5 hours per charge. With sleep modes and burst acquisition strategies not yet enabled in firmware, this represents a conservative lower bound on operational autonomy.

\section{Discussion and Conclusion}
This work presents a novel architecture for a self-calibrating, synchronized ultrasound microphone network using compact, wireless embedded nodes. By decoupling synchronization and data transmission across distinct radio bands—a Sub-1GHz link (868~\si{\mega\hertz})” and Wi-Fi 6 (2.4~\si{\giga\hertz} band) for high-throughput data—full wireless operation is achieved without sacrificing precision or reliability. Passive optical tracking via the HTC Vive Lighthouse system enables spatial self-localization without manual alignment.

Initial tests with a three-node setup confirmed core functionality. Each node sustained data rates of up to \SI{17}{\mega\bit\per\second} and maintained sub-microsecond inter-node synchronization. Battery-powered operation was determined to last up to \SI{5.5}{\hour} using a single-cell \SI{1400}{\milli\ampere\hour} LiPo battery, with further optimizations planned.

Future work will extend the system to support real-time streaming and sound source localization using distributed beamforming for applications such as animal vocalization~\cite{verreycken2021bio} studies and (industrial) acoustic monitoring~\cite{steckel2024toolwearpredictioncnc,8956631, MARTINEZ2023109618}. Overall, the prototype demonstrates a scalable, plug-and-play platform for advanced synchronized acoustic sensing in both scientific and industrial settings.

\clearpage
\newpage

\bibliographystyle{IEEEtran}
\bibliography{citationsLibrary}

\end{document}